\documentclass[%
 reprint,
superscriptaddress,
 amsmath,amssymb,
 aps,
]{revtex4-2}

\usepackage{graphicx}
\usepackage{dcolumn}
\usepackage{bm}

\usepackage{hyperref}
\usepackage[none]{hyphenat}
\usepackage{color}
\usepackage{rotating}

\begin{document}

\newcommand{\C}{$^\circ$C}
\newcommand{\muB}{$\mu_\mathrm{B}$}
\newcommand{\mueff}{$\mu_\mathrm{eff}$}
\newcommand{\TN}{$T_N$}
\newcommand{\DT}{$\Theta_\mathrm{D}$}
\newcommand{\ET}{$\Theta_\mathrm{E}$}

\preprint{APS/123-QED}

\title{Coexistence of Commensurate and Incommensurate Antiferromagnetic Groundstates in Co$_x$NbSe$_2$ Single Crystal}

\author{H. Cein Mandujano}
\affiliation{Department of Chemistry and Biochemistry, University of Maryland, College Park, Maryland 20742, USA}
\affiliation{Maryland Quantum Materials Center, Department of Physics, University of Maryland, College Park, Maryland 20742, USA}
 
\author{Peter Y. Zavalij}
\affiliation{Department of Chemistry and Biochemistry, University of Maryland, College Park, Maryland 20742, USA}

\author{Alicia Manj\'on-Sanz}
\affiliation{Neutron Scattering Division, Oak Ridge
National Laboratory, Oak Ridge, Tennessee 37831, USA}

\author{Huibo Cao}
\affiliation{Neutron Scattering Division, Oak Ridge
National Laboratory, Oak Ridge, Tennessee 37831, USA}

\author{Efrain E. Rodriguez}
\email{efrain@umd.edu}
\affiliation{Department of Chemistry and Biochemistry, University of Maryland, College Park, Maryland 20742, USA}
\affiliation{Maryland Quantum Materials Center, Department of Physics, University of Maryland, College Park, Maryland 20742, USA}


\date{\today}

\begin{abstract}


In Co$_x$NbSe$_2$, crystal symmetry, and cobalt site occupation drive the formation of two distinct magnetic phases. At $x = 1/4$, the centrosymmetric structure ($P$6$_3$/$mmc$) promotes Co-Co interactions leading to the formation of an $A$-type \color{black} antiferromagnetic structure \color{black} phase with a transition temperature of \TN$^A$ = 169 K. At $x = 1/3$, the non-centrosymmetric structure ($P$6$_3$22) induces a lower-temperature magnetic phase with \TN$^S$ = 28 K. We report the coexistence of both substructures within a superlattice, with a nuclear propagation vector of (1/3, 1/3, 0) relative to the host lattice.
Single crystals of Co$_{0.28}$NbSe$_2$ exhibit both magnetic transitions, with \TN$^A$ corresponding to the $x \sim 1/4$ phase and \TN$^S$ corresponding to the $x \sim 1/3$ phase. Magnetic susceptibility and specific heat measurements confirm these transitions, although only the high-temperature \TN$^A$ phase significantly affects resistivity. We successfully isolate each phase in powder samples, while single crystals with an intercalation ratio of $x = 0.28$ display the coexistence of both phases in a single sample.
Using single-crystal neutron diffraction, we solved the magnetic structure of the high-temperature centrosymmetric phase (\TN$^A$), and neutron powder diffraction revealed the double-$q$ magnetic structure of the low-temperature noncentrosymmetric phase (\TN$^S$).

\end{abstract}

\maketitle

 In transition metal dichalcogenides (TMDs) of the 2H-type, the electronic properties are intrinsically linked to the symmetry of the system. For the TMDs intercalated by magnetic 3$d$-metals, the concentration determines the structure and magnetic order. Depending on the intercalation level, these systems can adopt one of two distinct space groups, each associated with a specific phenomenon. When the intercalant concentration is $x \sim$ 1/4, the system exhibits a centrosymmetric space group $P$6$_3$/$mmc$. In this configuration, the intercalated $M$ ions form a 2$a_h\times$2$a_h$ sublattice, where $a_h$ is the unit cell length of the host $TMQ_2$ along the $a$-axis. This symmetry strongly influences the magnetic structure, leading to one set of magnetic correlations and higher transition temperatures (\TN) \cite{HULLIGER1970, Naik2022, Nakayama2006, Parkin1980_1, VOORHOEVE1970, VOORHOEVE1971, Xie2022, Hatanaka2023}. For ratios closer to $x \sim$ 1/3, the structure is non-centrosymmetric and chiral ($P$6$_3$22), with intercalated ions arranged in a $\sqrt{3}a_h \times \sqrt{3}a_h$ sublattice. This distinct symmetry leads to a different set of magnetic and structural properties, often associated with lower \TN\ values and different magnetic ordering \cite{Morosan2007, Mangelsen2020, Xie2022}. Furthermore, at the higher intercalation ratio limit, the magnetic interactions are extremely sensitive to the intercalant occupancy displaying an array of complex magnetic structures\cite{Park2024, Wu2022, Kousaka2022, Gubkin2016}. These well-defined symmetries play a critical role in determining the phase transitions and coexistence of multiple orders.  

The chemical composition or the ratio of intercalated magnetic ions to TMD host can be crucial in determining the structure and magnetic phase\cite{Hatanaka2023}.
In layered materials, phase separation due to stacking sequence, faults, or layer disorder is common, with layers accommodating various structural configurations\cite{Rodriguez2011, Vikas2011}. For instance, in CrI$_3$, three magnetic transitions coexist, defined by structural changes and phase separation \cite{Niu2020, Meseguer-Sanchez2021}. In such cases, each layer and its interaction with neighboring layers can be considered a single domain, creating a localized environment that affects the overall physical properties \cite{Cao2016, kim2024}. However, unexpected phenomena can emerge when structural discrepancies propagate between layers, as in interplanar domain ordering, further complicating the magnetic and structural behavior.

Coexisting order parameters, such as charge density waves (CDWs) and magnetic transitions, attract interest due to their potential applications in spintronics, particularly properties of exchange bias \cite{Wadley2016, Jungwirth2016}. In TMDs intercalated with magnetic ions, the interaction between these coexisting phases and the material’s ground state, defined by site occupation and antiferromagnetic spin configuration, can lead to exchange bias and other novel properties \cite{Maniv2021, Maniv2021b}.

A canonical example of a TMD hosting both superconductivity and a CDW is pristine 2H-NbSe$_2$, which undergoes a superconducting transition at 7 K and a CDW transition at 31 K \cite{Majumdar2020, Mallikas2013, Iavarone2008}. Intercalating 3$d$ metal ions into the van der Waals (vdW) gaps of NbSe$_2$ has further expanded the plethora of magnetic phenomena, offering new platforms for studying phase competition and coexistence \cite{Naik2022, Iavarone2008, Hauser1973}.

\begin{figure}[t]
\includegraphics[scale=0.40]{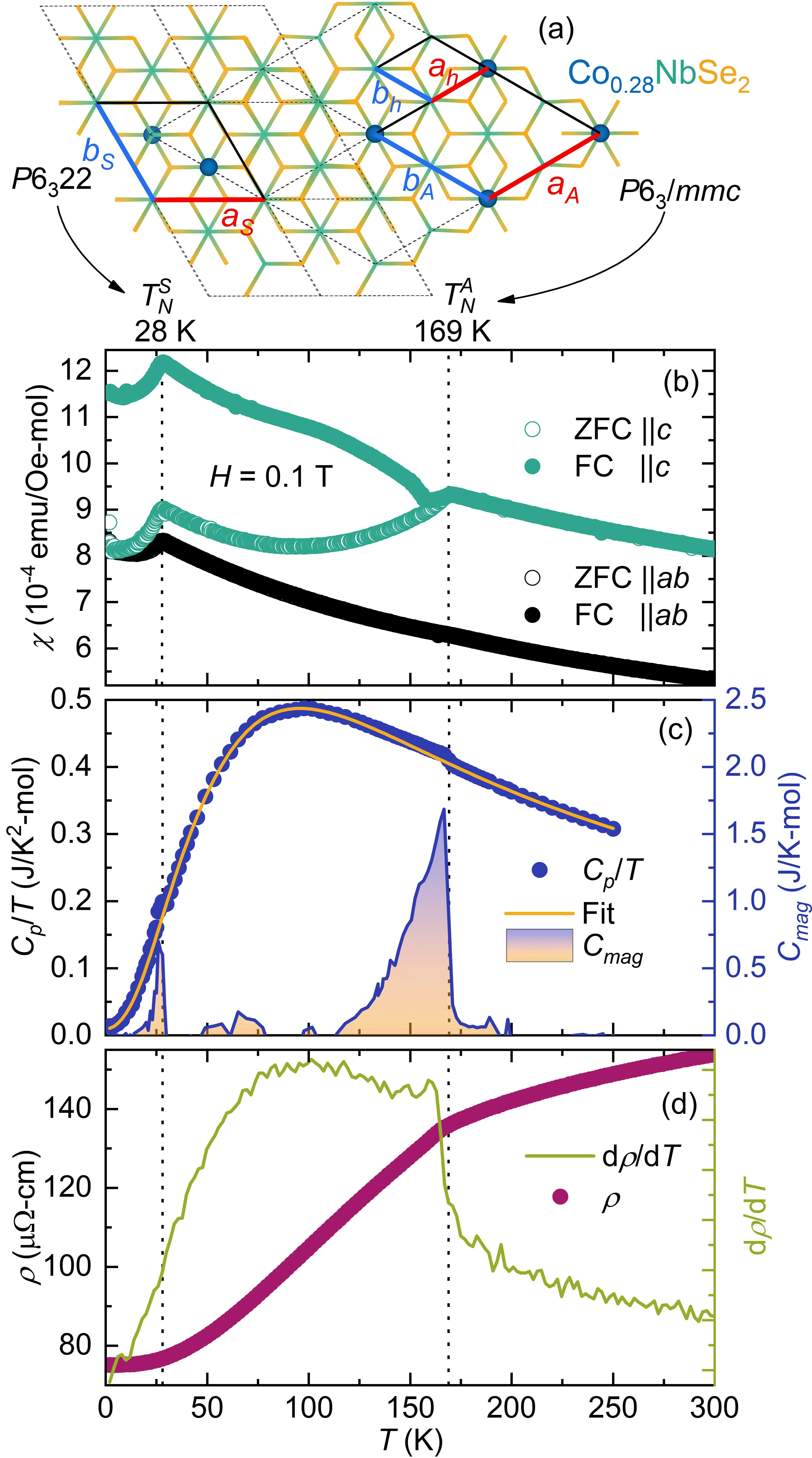}
\caption{\label{fig_properties} Structure association to bulk properties of Co$_{0.28}$NbSe$_2$. (a) Layered 2H-polytype of NbSe$_2$ host lattice viewed along the $c$-axis with $a_h\approx$ 3.44 \AA. On the left is the noncentrosymmetric $P$6$_3$22 where $a_S$ = $\sqrt{3}a_h\times\sqrt{3}a_h$. To the right is the centrosymmetric $P$6$_3$/$mmc$ where $a_A$ = 2$a_h\times$2$a_h$. These cobalt arrangements originate the distinct magnetic phases observed in the directional zero field-cooled and field-cooled magnetic susceptibility (b) with field applied out of plane (green) and in-plane (black).
(c) Specific heat over temperature (blue) and Debye-Einstein fit (yellow); the subtraction provides the net magnetic specific heat.
(d) Resistivity (red) and its derivative (light green).}
\end{figure}

In this work, we revisit Co$_x$NbSe$_2$, previously studied by Voorhoeve et al. \cite{VOORHOEVE1971}, and \color{black} is recently being proposed as an $A$-type altermagnet for $x$ = 1/4  by Regmi et al\cite{Regmi2024, Dale2024}. \color{black}Our single-crystal studies focus on Co$_{0.28}$NbSe$_2$ containing both sublattices, from which we exclusively observe the magnetic phase associated with the $x \sim 1/4$ ratio through single-crystal neutron diffraction. We provide novel insights into the magnetic structure of Co$_{1/3}$NbSe$_2$ for the first time using neutron powder diffraction. Through these structural measurements, we explore these magnetic transitions and their coexistence, deriving the magnetic structures using irreducible representation analysis.

We synthesized powder samples of Co$_{x}$NbSe$_2$ with nominal compositions of $x$ = 1/3 and $x$ = 1/4 by mixing elemental powders in molar ratios of 1.15:3:6 and 1:4:8, respectively. The mixed powders were vacuum-sealed in fused silica tubes and annealed at 900 \C\ for 5 days.
Phase formation and purity were verified via the Rietveld method on the powder X-ray diffraction using GSAS-II\cite{gsas}.
\color{black}For crystal growth, we aimed to obtain Co$_{1/3}$NbSe$_2$ via chemical vapor transport (CVT), using 3.5 mg/mL of I$_2$ as the transport agent and the corresponding powder as the charge source. The CVT was carried out in a 12.5 cm long ampule with an inner diameter of 14 mm. The vacuum-sealed ampule was placed in a single-zone furnace with a natural temperature gradient of 80 \C; the heating profile was a one-step heating ramp to 875 \C\ at the hot zone and 815 \C\ at the cold growth end. This temperature was maintained for 10 days before allowing the furnace to naturally cool. \color{black}Despite compensating for cobalt loss due to the formation of CoI$_2$ by adding excess cobalt powder during vapor transport, the resulting composition consistently yielded Co$_{0.28\pm\delta}$NbSe$_2$.
We verified the composition of the crystals using single-crystal X-ray diffraction (XRD) and energy-dispersive X-ray spectroscopy (EDS). Magnetic susceptibility and bulk property measurements were performed using a Quantum Design magnetic properties measuring system and a Dynacool physical property measuring system, respectively. \color{black}Electrical transport was done using 4-probes\color{black}. The nuclear structure was determined through single-crystal X-ray diffraction with SHELXT program and refined in the SHELXL\cite{Krause2015, Sheldrick2015}.
Single-crystal neutron diffraction experiments on Co$_{0.28}$NbSe$_2$ were conducted at the DEMAND (HB-3A) beamline with $\lambda = 1.546$ \AA\ at the High Flux Isotope Reactor (HFIR) at Oak Ridge National Laboratory (ORNL)\cite{DEMAND2019}. Neutron powder diffraction experiments for Co$_{x}$NbSe$_2$ with $x$ = 1/3 and $x$ = 1/4 were performed at the POWGEN (BL-11A) beamline with $\lambda = 1.5$ \AA\ at the Spallation Neutron Source (SNS) at ORNL \cite{POWGEN_Huq2011}.

\begin{figure}[t]
\includegraphics[scale=0.52]{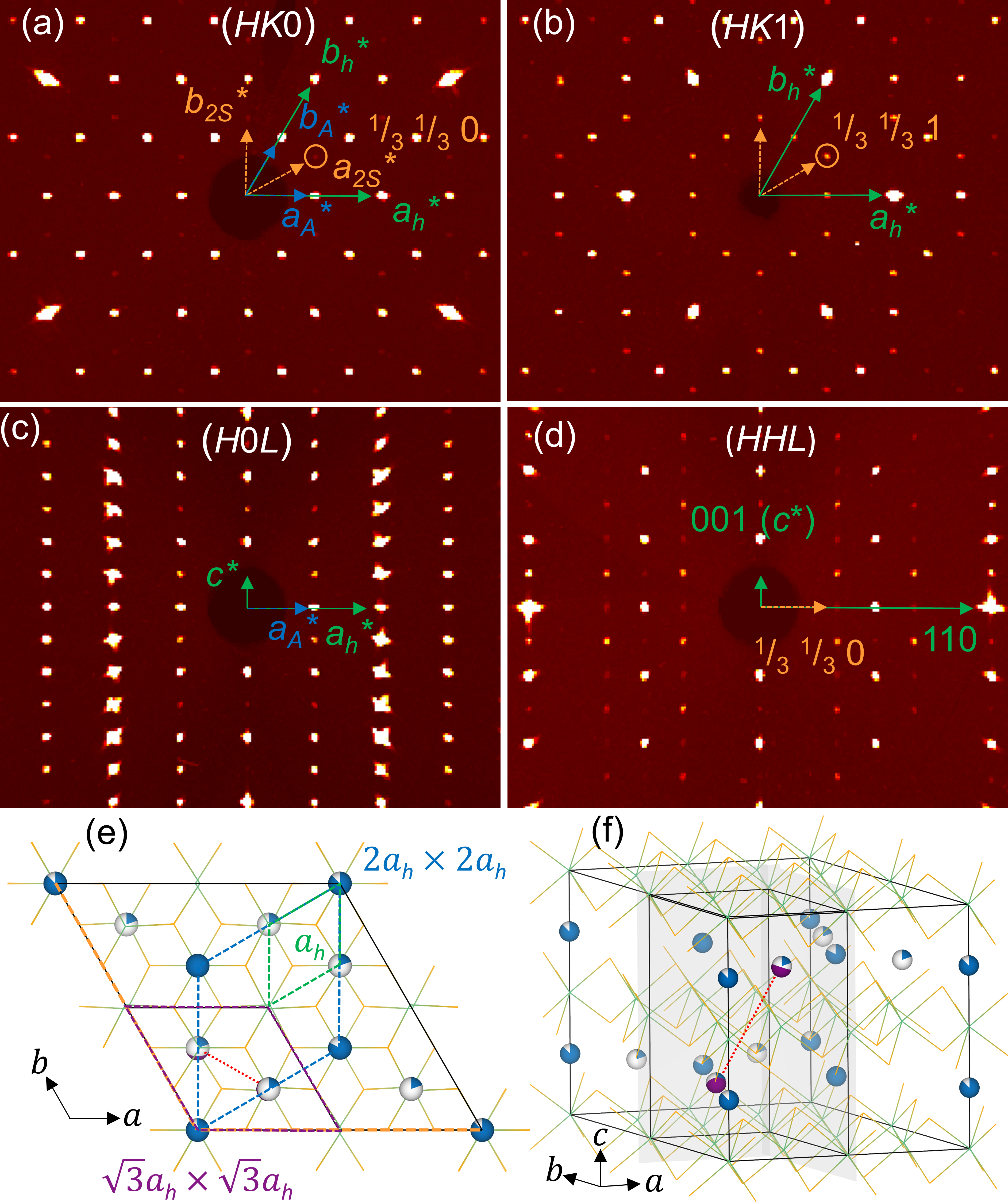}
\caption{\label{fig_mod} Single crystal X-ray diffraction results showing lattice coexistence. Precession images along the ($HK$0) (a), ($HK$1) (b), ($H$0$L$) (c), and ($HHL$) (d) planes. Green arrows indicate the host unit cell of 2H-NbSe$_2$ and the yellow vector points toward the observed satellite reflections indexed by the nuclear propagation vector $q_{2S}$ = (1/$3$, 1/3, 0). The blue arrows correspond to the 2$a_h\times$2$a_h$ cell ($a_A$). The resulting structure solution with 2$\sqrt{3}a_h$ $\times$ 2$\sqrt{3}a_h$ ($a_{2S}$) dimensions viewed along the $c$-diretion (e) and $ab$-plane (f). The red dotted line marks the trajectory to the next-nearest neighboring cobalt atom in the adjacent layer, defining the $\sqrt{3}a_h \times \sqrt{3}a_h$ domains (labeled as $a_S$). }
\end{figure}

\begin{figure}[t]
\includegraphics[scale=0.37]{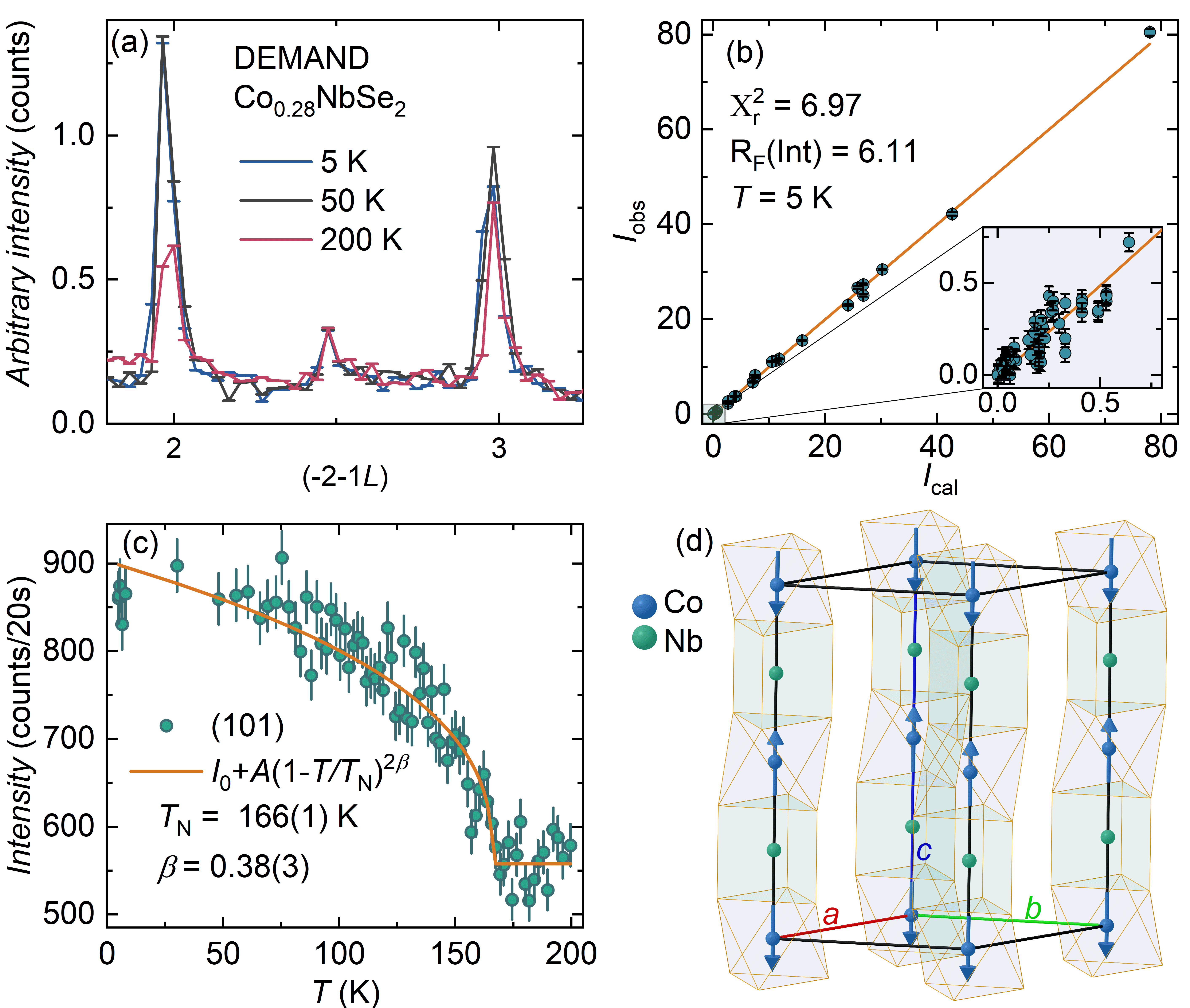}
\caption{\label{fig_DEMAND}  Single crystal neutron diffraction results. (a) Intensity of Co$_{0.28}$NbSe$_2$ along (-2 -1 $L$).
The peak at (-2 -1 2.5) corresponds to $\lambda$/2.
(b) Observed versus calculated neutron diffraction intensities.
(c) The magnetic order parameter collected upon the (101) reflection and is fitted to the power law function.
(d) Magnetic structure solution. 
The nuclear structure was solved via single crystal XRD and was used to obtain the magnetic structure using representation analysis in Mag2Pol\cite{Mag2Pol_Qureshi2019}.}
\end{figure}

 \begin{figure*}[!t]
 \includegraphics[scale=0.62]{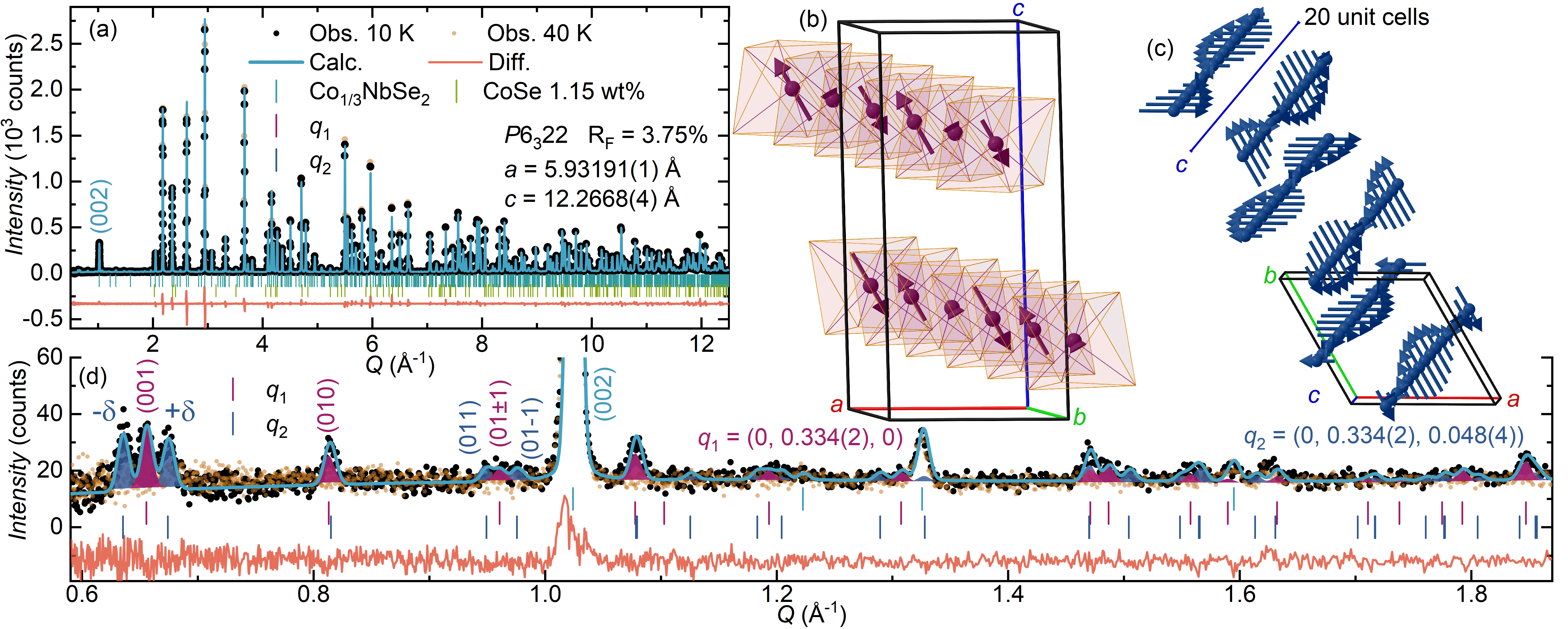}
 \caption{\label{fig_NPD} Neutron powder diffraction of Co$_{1/3}$NbSe$_2$. Rietveld refinement results at (a) 10 K overlayed on 40 K. 
 Resulting magnetic structure obtained from commensurate $q_1$ (b) and incommensurate $q_2$ (c).
 (d) Low-$Q$ region of (a) showing the two magnetic propagating vectors; $q_1$ (red) and $q_2$ (blue).}
 \end{figure*}

The physical properties of Co$_{0.28}$NbSe$_2$, as shown in Figure \ref{fig_properties}(b-d), reveal two distinct magnetic phase transitions coexisting within a single crystal sample.
Powder samples, however, indicate that each transition corresponds to a defined symmetry inherent to the intercalated cobalt ratio (refer to supplemental material\cite{SI}).
As illustrated by Figure \ref{fig_properties}(a), the low transition temperature (\TN$^S$) is associated with the non-centrosymmetric $P$6$_3$22 (No. 182) symmetry while the high transition temperature (\TN$^A$) pertains to the higher symmetry and centrosymmetric $P$6$_3$/$mmc$ (No. 194).
The \TN$^A$ transition is observable only in the $\chi||c$ configuration (Figure \ref{fig_properties}(b)), indicating a strong anisotropy and defining the antiferromagnetic easy axis for this phase.
In the case of \TN$^S$, the transition is visible in both applied field directions, indicating distinct magnetic interactions.

We estimate the magnetic entropy associated with each phase by subtracting the Debye-Einstein fit from the experimental specific heat data (Figure \ref{fig_properties}(c)). The larger area under the \TN$^A$ peak indicates stronger magnetic correlations in this phase. Furthermore, only \TN$^A$ appears to influence the Fermi surface, as demonstrated by the temperature-dependent resistivity measurements. The impact is made clear through the first derivative of the resistivity (Figure \ref{fig_properties}(d)).
The lack of a discernible response from \TN$^S$ in resistivity does not necessarily imply that the Fermi surface remains unchanged. The sulfur analog, Co$_{1/3}$NbS$_2$ for example, has been shown to host nontrivial 2D bands in the antiferromagnetic state, observed via anomalous Hall measurements\cite{Ghimire2018, Tenasini2020}.
We speculate that the absence of a response at \TN$^S$ could be attributed to the dilute magnetic scattering of the $P$6$_3$22 domains. This aligns with observations in other intercalated systems, such as Li$_x$Fe$_{1/4}$NbS$_2$, where the $\sqrt{3}a\times\sqrt{3}a$ motif domain phases are suppressed with increasing lithiation\cite{Lawrence2023}.
\color{black}
To the best of our knowledge, while structural phase coexistence in 3$d$-metal intercalated TMDs has also been reported recently by Xie et al.\cite{xie2024}, our work demonstrates well-defined magnetic transitions that can be distinctly associated with each structural phase, providing new insights into their interplay.
\color{black}

From Figure \ref{fig_properties}(a), it is clear how the rigid 2H-NbSe$_2$ framework can accommodate intercalated cobalt sites generating two distinct symmetries.
Having identified both high- and low-temperature magnetic transitions, we examined the structural coexistence of these phases using single-crystal XRD.
As a first approach, we solve the structure of Co$_{0.28}$NbSe$_2$ in the $P$6$_3$/$mmc$ spacegroup, with the characteristic 2$a_h\times$2$a_h$ ($a_A$) sublattice and unit cell dimensions of $a_A$ = 6.90580(10) \AA\ and $c$ = 12.3630(4) \AA.
However, upon a closer inspection of the precession images (Figure \ref{fig_mod}(a-d)) we observe satellite reflections.
These satellite reflections are indexed by the nuclear propagating vector $q_{2S}$ = (1/3, 1/3, 0) with respect to the host unit cell. These satellite reflections can equivalently be indexed by $q_{2S}$ = (2/3, 2/3, 0) on the basis of the 2$a_h\times$2$a_h$ lattice.
This larger supercell rendered in Figure \ref{fig_mod}(e) is solved in the $P$6$_3$22 space group, with twice the lattice dimensions of the $\sqrt{3}a_h\times\sqrt{3}a_h$ cell ($a_{2S}$ = 11.9622(3) \AA).
This supercell allocates cobalt sites such that both the $a_A$ (2$a_h\times$2$a_h$) and the $a_S$ ($\sqrt{3}a_h\times\sqrt{3}a_h$) sublattices coexist (Figure \ref{fig_mod}(e and f)).
The Miller indices of the $a_{2S}$ supercell and the host lattice are related by:

\begin{center}
$\begin{bmatrix}
h\\
k\\ 
l\\
\end{bmatrix}^{a_h}$
=
$\begin{bmatrix}
1/6 & 1/6 & 0\\
-1/6 & 1/3 & 0\\
0 & 0 & 1
\end{bmatrix}$
$\begin{bmatrix}
h\\
k\\ 
l\\
\end{bmatrix}^{a_{2S}}$
\end{center}

Where the superscripts refer to the corresponding lattice.
The site occupancy and their intralayer distances significantly influence their magnetic interactions. The higher cobalt intercalation ratio disrupts spatial inversion symmetry, leading to two distinct magnetic ground states \cite{Zheng2021, Xie2022, Parkin1980_1}.
Despite our results, domains may not necessarily coexist in the same layer but stacked; some layers containing 2$a_h\times$2$a_h$ domains and others with $\sqrt{3}a_h\times\sqrt{3}a_h$ appearing as both phases being superimposed.

Provided with a crystal containing both magnetic phases, we first studied Co$_{0.28}$NbSe$_2$ single crystal sample with neutron diffraction to obtain the magnetic structure.
However, no satellite reflections were observed.
Upon closer examination of the nuclear reflections, we noted that certain Bragg intensities increased below \TN$^A$, as demonstrated by the (-2 -1 2) reflection (Figure \ref{fig_DEMAND}(a)). This intensity growth observed between 200 K and 50 K, is consistent with the \TN$^A$ transition, but no further changes were detected down to 5 K. Therefore, all magnetic reflections correspond exclusively to \TN$^A$.
The magnetic reflections overlapping with the nuclear ones indicate a propagating vector $q_m$ = (0 0 0).
Through irreducible representation analysis, we identified twelve possible magnetic structures. 
Based on susceptibility data, we inferred that the magnetic moment aligns along the $c$-axis. Only two representations, $\Gamma_3$ (ferromagnetic) and $\Gamma_7$ (antiferromagnetic), meet this requirement. Ultimately, we deduced the correct magnetic structure, corresponding to the antiferromagnetic $\Gamma_7$ representation (Figure \ref{fig_DEMAND}(b) and (d)).
The resulting structure is equivalent to the $P$6$_\mathrm{3}'\slash m'm'c$ magnetic space group that describes other $M_{1/4}TQ_2$ $A$-type structures\cite{Lawrence2023, Vanlaar1971, Mandujano2024}  and recently reported in Co$_{1/4}$NbSe$_2$\cite{Regmi2024}. This structure and composition have attracted attention for its spin and spatial symmetry elements identifying it as a potential altermagnet\cite{Smejkal2022, Smejkal20222}.
We find a total magnetic moment of 1.5(4) \muB/Co$^{2+}$, which is lower than  expected for $S$ = 3/2.
The depleted moment likely arises from the interaction between the intercalant’s 3$d$ orbitals and the host material's $d_{z^2}$ itinerant electrons from vertically neighboring Nb \cite{Motizuki1992, Naik2022, Maksimovic2022, Mandujano2024}.

We follow the temperature evolution of \TN$^A$ ordering by measuring the intensity of the (101) nuclear Bragg reflection as a function of temperature (Figure \ref{fig_DEMAND}(c)).
We find that the integrated peak intensity of the nuclear reflection increases near the high-temperature magnetic anomaly.
To obtain the critical exponent $\beta$, we fit the order parameter data to the power law expressed by $I$ = $I_0$+$A$(1-$T/T_N)^{2\beta}$.
The obtained critical value $\beta$ = 0.38(3) best approximates the Heisenberg class ($\beta$ = 0.366)\cite{Pelissetto2002}.
Given the dimensionality of the structure and $A$-type magnetic ordering, it would be intuitive to assume the exponent to behave as Ising class\cite{Ko2011, Wu2023}.

Given the absence of magnetic reflections from the \TN$^S$ phase in the single-crystal neutron diffraction, we turned to neutron powder diffraction to investigate Co$_{1/3}$NbSe$_2$. Powder diffraction data, as shown in Figure \ref{fig_NPD}(a), refined to the non-centrosymmetric $P$6$_3$22 space group at 40 K, revealing purely nuclear reflections. When comparing data collected at 40 K and 10 K, we observed the appearance of magnetic peaks associated with the \TN$^S$ transition. The powder sample exhibited a single magnetic phase, corresponding to the low-temperature \TN$^S$ state.
At low-$Q$ values (Figure \ref{fig_NPD}(d)), we identified a distinct triplet-like feature, suggesting the coexistence of two magnetic propagating vectors: a commensurate vector, $q_1$ = (0, 1/3, 0), and an incommensurate vector, $q_2$ = (0, 1/3, 0.048(4)). The middle peak in the triplet corresponds to the (001) reflection, indexed by $q_1$, while the side peaks arise from incommensurate modulations of $q_2$.
Refining the magnetic structure, we find $q_1$ has a small in-plane component. Still, most of its magnetic moment is along the $c$-direction with a total magnetic moment of 1.2(2)\muB/Co$^{2+}$ refined using $\Gamma_2$ with equal opposing magnitude constraints (Figure \ref{fig_NPD}(b)).
The magnetic order from $q_2$ fits to a sine wave shape oscillating along the $c$-axis with 2.9(2)\muB/Co$^{2+}$ total moment, which is close to the expected value. The length of this propagating wave is approximately 255 \AA spanning 20 unit cells (Figure \ref{fig_NPD}(c)).
The individual moments composing this sine wave are parallel to the $ab$-plane. The fixture of the incommensurate component of $q_2$ remains an open question as it may undergo a commensurate locking at a lower temperature\cite{Moncton1975}.
The oscillating spin moments observed here corroborate the presence of Dzyaloshinskii-Moriya (DM) interaction when $x\geq$ 1/3 where spatial inversion symmetry is broken\cite{An2023, Wu2023, Xie2022}.
Additionally, the RKKY mechanism is inversely dependent on 
$R_{ij}^3$ (where $R$ is the distance between metallic centers), may be used to understand the preferred moment orientation of both AFM orders encountered\cite{Ruderman1954, Kasuya1956, Yosida1957}.

Our study indicates that the intercalation of Co ions into pristine 2H-NbSe$_2$ suppresses both the CDW and superconductivity. However, CDW has been observed to coexist with magnetic ordering in Fe$_x$NbS$_2$ \cite{Wu2023}. In the novel Co$_x$NbSe$_2$ phases investigated here, similar charge instabilities may enhance exchange interactions between the intercalated localized magnetic moments.

We successfully isolated Co$_x$NbSe$_2$ polycrystalline phases for $x = 1/4$ and $1/3$ while assigning their magnetic structures. We propose that the simplicity of the \color{black} collinearly compensated $A$-type order \color{black} in the $x = 1/4$ phase arises from the vertically ordered Co sites and their alignment with the Nb 4$d{z^2}$ orbital. In contrast, the more complex double-$q$ magnetic structure observed in the $x = 1/3$ phase likely contributes to enhanced metallic interactions, including DM interactions.
Both structural and magnetic phases coexist within a 2$\sqrt{3}a_h \times$2$\sqrt{3}a_h$ superlattice, accommodating these distinct magnetic orders. To fully address the implications between neighboring magnetic centers, future inelastic neutron scattering experiments are essential for exploring this rich compound series.

\section{Acknowledgments}
The authors thank the U.S. Department of Energy
(DOE), Office of Science (Grant DE-SC0016434), for financial support.
Use of the High Flux Isotope Reactor (DEMAND experiment IPTS-31569) and Spallation Neutron Source (POWGEN experiment IPTS-31862) at Oak Ridge National Laboratory was supported by the U. S. Department of Energy, Office of Science, Office 
of Basic Energy Sciences, under Contract No. DE-AC02-06CH11357.
We also acknowledge support from the Quantum Materials Center. 
This work has been supported in part by the X-ray Crystallographic Center at 
The University of Maryland.


\bibliography{Co13n14NbSe2_Ref}

\pagebreak
\newpage

\end{document}